%% file: icrc2021-cta-pipelines-proceeding.tex
\documentclass[a4paper,11pt]{article}
\usepackage{pos}
\usepackage[super]{nth}

\usepackage{graphicx}
\usepackage{url}
\graphicspath{ {./} }

\usepackage[american]{babel}
\usepackage[autostyle]{csquotes}
\usepackage[backend=biber, giveninits=true]{biblatex}

\usepackage[locale=US]{siunitx}
\usepackage{ragged2e}
\usepackage[para]{footmisc}
\usepackage{enumitem}

\usepackage{caption}
\usepackage{subcaption}

\usepackage{bookmark}

\DeclareSourcemap{
 \maps[datatype=bibtex,overwrite=true]{
  \map{
    \step[fieldsource=Collaboration, final=true]
    \step[fieldset=usera, origfieldval, final=true]
  }
 }
}

\renewbibmacro*{author}{%
  \iffieldundef{usera}{%
    \printnames{author}%
  }{%
    \printnames{author} for the \printfield{usera}%
  }%
}%

\title{Prototype Open Event Reconstruction Pipeline for the Cherenkov Telescope Array}
 \ShortTitle{Prototype Open Event Reconstruction Pipeline for CTA}

\manuallySeparateAuthors
\author*[a]{M.\ Nöthe}
\author[b]{, K.\ Kosack}
\author[a]{, L.\ Nickel}
\author[b]{ and M.\ Peresano}
\author{ for the CTA Consortium}
\renewcommand{\printHeadAuthors}{M.~Nöthe et al. for the CTA Consortium}

\affiliation[a]{TU Dortmund University,\\
  Otto-Hahn-Str. 4a, Dortmund, Germany}

\affiliation[b]{AIM, CEA, CNRS, Universite Paris-Saclay, Universite Paris Diderot, Sorbonne Paris Cite, F-91191 Gif-sur-Yvette, France}

\emailAdd{maximilian.noethe@tu-dortmund.de}

\newcommand\shorturl[1]{\href{https://#1}{\texttt{\detokenize{#1}}}}

\abstract{%
    The Cherenkov Telescope Array (CTA) is the next-generation gamma-ray observatory
    currently under construction.
    It will improve over the current generation of imaging atmospheric Cherenkov telescopes (IACTs)
    by a factor of five to ten in sensitivity and it will be able to observe the whole sky from a combination of two sites:
    a northern site in La Palma, Spain, and a southern one in Paranal, Chile.
    CTA will also be the first open gamma-ray observatory.
    Accordingly, the data analysis pipeline is developed as open-source software.
    The event reconstruction pipeline accepts raw data of the telescopes and processes it to
    produce suitable input for the higher-level science tools.
    Its primary tasks include reconstructing the physical properties of each recorded
    shower and providing the corresponding instrument response functions.

    \texttt{ctapipe} is a framework providing algorithms and tools to facilitate raw data calibration,
    image extraction, image parameterization and event reconstruction.
    Its main focus is currently the analysis of simulated data but it has also been successfully applied
    for the analysis of data obtained with the first CTA prototype telescopes, such as the Large-Sized Telescope 1 (LST-1).

    \texttt{pyirf} is a library to calculate IACT instrument response functions,
    needed to obtain physics results like spectra and light curves,
    from the reconstructed event lists.

    Building on these two, \texttt{protopipe} is a prototype for the event reconstruction pipeline for CTA.
    Recent developments in these software packages will be presented.
}

\FullConference{\nth{37} International Cosmic Ray Conference (ICRC 2021)\\
		July \nth{12} -- \nth{23}, 2021\\
		Online -- Berlin, Germany}

\begin{document}
\maketitle

\begin{figure}[t]%
  \begin{subfigure}{0.25\textwidth}
    \includegraphics{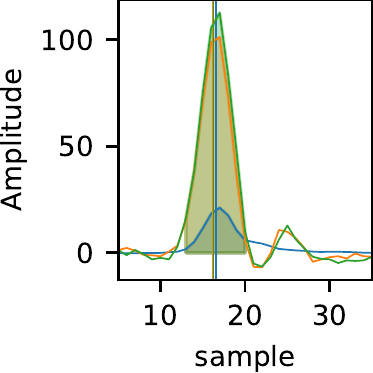}
  \end{subfigure}%
  \begin{subfigure}{0.5\textwidth}
    \includegraphics{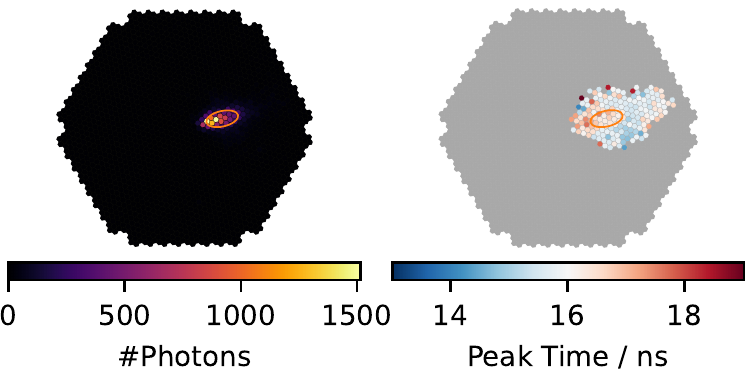}
  \end{subfigure}%
  \begin{subfigure}{0.25\textwidth}
    \includegraphics{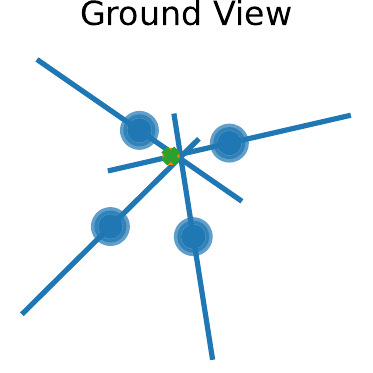}
  \end{subfigure}%
  \caption{Steps of the analysis up to shower geometry reconstruction: 1.\ The Cherenkov pulses in each pixel are found and integrated (left) to obtain the number of photons (second from left) and peak times (third from left). 2.\ The resulting image is cleaned, the pixels not selected are shown in gray in the peak time plot. 3.\ The images are parametrized, including the Hillas parameters which are visualized using an ellipse. 4.\ The physical properties of the primary are reconstructed. The plot on the right shows the impact point of the primary on the ground.}\label{fig:steps}
\end{figure}

\section{Introduction}

\noindent The Cherenkov Telescope Array (CTA)\footnote{\shorturl{www.cta-observatory.org}}
will be the next generation very-high-energy gamma-ray observatory, sensitive to energies between $\sim\SI{20}{\GeV}$ and  \SI{300}{\TeV}.
It will be composed of over fifty imaging atmospheric Cherenkov telescopes (IACTs) built at two sites to achieve full sky coverage:
one on the Canary Island of La Palma, Spain and the other near Paranal, Chile.

CTA will detect gamma rays by measuring the Cherenkov light emitted by extensive air showers,
however these are also induced by charged cosmic rays, which form a large background to gamma-ray observations.
The data analysis pipeline of CTA starts with the pre-calibrated raw data from the telescopes in the form of
time series data for each pixel and for each telescope that registered a signal from the current shower.
The pipeline proceeds to reconstruct the physical properties of the primary particle for each recorded shower,
this includes the gamma ray's energy and arrival direction.
To remove most of the cosmic-ray induced air showers, also a particle type classification is required.

In the classical analysis approach, the raw data is reduced and aggregated into higher levels of abstraction before finally employing a set of machine learning and geometrical algorithms to reconstruct the physical properties of the primary particle.
This is usually performed as a four step procedure, which is currently implemented in \texttt{ctapipe}:
image extraction, image cleaning, image parametrization and finally the reconstruction of primary particle properties (see Figure \ref{fig:steps}).
These steps will be detailed in \autoref{sec:ctapipe}.
After reconstruction of the shower events, one last step – based on the \texttt{pyirf} library described in \autoref{sec:pyirf} – selects the best ones on the basis of the specific science case at hand and allows to produce the instrument response functions (IRFs).
The pipeline prototype called \texttt{protopipe} and described in section \autoref{sec:protopipe} performs the analysis steps from raw simulated data to IRF production based on both libraries.

\section{\texttt{ctapipe}}\label{sec:ctapipe}\noindent
\texttt{ctapipe} is a python package providing library functions and command-line tools to perform the tasks listed in the previous section.
It is developed as open-source software and the project, started in 2015, is hosted on Github.
Since then, 26 versions have been released but it is still under heavy development (the latest release at the time of writing is 0.11.0 ~\cite{ctapipe}).
In total, 44 contributors have made this project possible.
Releases are published to PyPI and conda packages are provided using \texttt{conda-forge}\footnote{\shorturl{conda-forge.org}}.
\texttt{ctapipe} builds upon the scientific python stack with the main dependencies being \texttt{astropy}~\cite{astropy} for astronomical computations and unit support, \texttt{numpy}~\cite{numpy} and \texttt{scipy}~\cite{scipy} for numerical algorithms and statistics and \texttt{pytables}\footnote{\shorturl{www.pytables.org}} for IO using HDF5\footnote{\shorturl{www.hdfgroup.org/solutions/hdf5/}}.
The jit-compiler \texttt{numba}~\cite{numba} is used to optimize performance-critical parts of the code base.

\subsection{Image Extraction}\label{image_extraction}\noindent
The first step in the \texttt{ctapipe} analysis is to reduce the time-series information,
i.\,e.\ the digitized signals of the Cherenkov photosensors,
to the number of photons and their mean arrival time in each pixel.
\texttt{ctapipe} supports different algorithms for extracting these quantities from single-pixels waveforms,
from simple peak finding algorithms to more complex ones which combine the waveforms of multiple pixels or
that fit the expected time evolution of the shower and use that to define the integration window for each pixel.

\subsection{Image Cleaning}\noindent
This operation is aimed at identifying pixels which are likely to host real Cherenkov signal.
This is usually done by applying a pixel-wise selection via cleaning thresholds based on the photo-electron
and peak time values output by the image extraction step.
Again, \texttt{ctapipe} supports multiple algorithms to solve this task.

\subsection{Image Parametrization}\label{image_parametrization}\noindent
After removal of noise pixels, the cleaned image goes through a parametrization in order to make it exploitable by subsequent algorithms, in particular shower geometry reconstructors and/or machine-learning models that assist with with the 
event property reconstruction.
Among the most important parameters are the classical Hillas parameters~\cite{hillas}, which describe
the orientation and extension of the shower image in the camera, which is needed for the following reconstruction steps.
Additionally, \texttt{ctapipe} implements general descriptive statistics of the images,
morphological features like the number of isolated pixel groups and
parameters describing the containment of the shower's image in each camera.

\subsection{Reconstruction of Event Properties}\label{estimation_events}\noindent
While the first three steps can be performed individually for each telescope in the array (monoscopic),
this step needs to combine the information from all telescopes to give one common estimate for a recorded shower (stereoscopic).

The stereoscopic reconstruction of physical shower parameters can be performed in \texttt{ctapipe} by either of two currently supported approaches: moments-based and template-based.

The moments-based method makes use of a reconstructor which takes as input the parametrized moments of each image (in the default approach the Hillas parameters) from a candidate shower. This input is then combined with a pair-wise geometric reconstruction where each pair of images gets a weight based on the brightness of the images.
In case of single, monoscopic telescopes, machine learning can also be used for the reconstruction of the origin, as the geometrical approaches require multiple telescopes.

\texttt{ctapipe} also supports the \texttt{ImPACT}~\cite{impact} algorithm, an advanced template-based likelihood optimization to reconstruct the event properties, where the expected image for a given set of event properties is calculated
from simulations, stored in a database of template images which is then used to perform a likelihood fit to the observed image.

\subsection{Input / Output, visualization and configuration}\noindent
IACT events are read from input files using the \texttt{EventSource} interface,
which can be implemented for custom file formats using the \texttt{ctapipe} plugin system\footnote{\RaggedRight E.\,g.: \shorturl{github.com/cta-observatory/ctapipe_io_lst}}.
There are built-in event source implementations for the simulation file format and \texttt{ctapipe}'s own output data format.
\texttt{ctapipe}'s data model uses its own data structure, called \texttt{Container}, which can be written to and loaded
from HDF5 files, supporting transformations and metadata including units.

The \texttt{ctapipe.visualization} module provides classes to display both camera images and telescope array configurations.
Two implementations currently exist, one using \texttt{matplotlib}~\cite{matplotlib} and one using the \texttt{bokeh}\footnote{\shorturl{bokeh.org/}} library.

The \texttt{ctapipe} configuration system is build using traitlets\footnote{\shorturl{traitlets.readthedocs.io/}},
the configuration system developed for IPython.
A full configuration tree is built by configurable classes called \texttt{Components} that can include
configurable member attributes.
Command-line tools use the same configuration mechanism and allow passing configuration for all
configurable objects either on the command line or via a configuration file.
Many options can be set per telescope type or even per telescope.

\section{Calculating Instrument Response Functions (IRFs) using \texttt{pyirf}}
\label{sec:pyirf}
\noindent To be able to estimate physical properties of gamma-ray sources from lists of reconstructed events, the instrumental response to the initial gamma-ray signal must be known.
This will depend on the instrument, the specific analysis, environmental conditions and more.
In general, the instrumental response of a gamma-ray telescope can be described by the following integral equation, transforming true properties of the gamma rays into the observable quantities:

\begin{equation}
    e(\hat{\alpha},\hat{\delta},\hat{E},t)
    = \int R(\hat{\alpha},\hat{\delta},\hat{E}|\alpha,\delta,E,t)
    \cdot I(\alpha,\delta,E,t)
    \, \mathrm{d}\Omega \, \mathrm{d}E
     + b(\hat{\alpha}, \hat{\delta}, \hat{E})
    \label{eq:irf}
\end{equation}
Where $\alpha$, $\delta$ and $E$ are the right ascension, declination of the gamma ray origin and its total energy, while $\hat{\alpha}$, $\hat{\delta}$, $\hat{E}$ are the corresponding reconstructed quantities obtained from the analysis pipeline. $I$ is the source term, the true gamma-ray signal arriving at earth at the given position, energy and time $t$. $R$ is the IRF, the convolution kernel translating true quantities to the observed ones, $b$ is the irreducible background and $e$ is the expected event distribution as measured by the experiment.
The solid angle integration over $\alpha, \delta$ is denoted using $\mathrm{d}\Omega$.

The IRF can only be estimated from labeled data, where the true and reconstructed quantities are both known.
In the case of CTA these labeled datasets are created via Monte Carlo simulations using CORSIKA~\cite{corsika} to simulate the extensive air showers, followed by the detector simulation performed by \texttt{sim\_telarray}~\cite{simtel}.

In classical IACT analysis, the IRF is factorized into three independent components, making the strong assumption that the migrations between the different observables are statistically independent.
This factorization yields:
\begin{equation}
    R(\hat{\alpha},\hat{\delta},\hat{E}|\alpha,\delta,E,t) =
    A_\text{eff}(\alpha,\delta,E,t) \cdot
    \operatorname{PSF}(\hat{\alpha}, \hat{\delta} | \alpha,\delta,E,t) \cdot
    D(\hat{E} | \alpha,\delta,E,t)
    \label{eq:factorizedirf}
\end{equation}
Where $A_\text{eff}$ is the effective area, the detection probability times the observed area for a gamma ray with given true properties, $\operatorname{PSF}$ is the point spread function, i.\,e.\ the convolution kernel for the reconstructed gamma-ray origin and $D$ is the energy dispersion, the migration between true energy $E$ and reconstructed energy $\hat{E}$.
Instead of continuous functions, these IRFs are calculated and stored as binned quantities filled from simulated events.

\texttt{pyirf} is a python library for calculating these IRFs from labeled, reconstructed event lists as created by the event reconstruction pipeline.
The latest version of \texttt{pyirf} at the time of writing is \texttt{v0.5.0}~\cite{pyirf}, which supports calculating most IRFs formats defined in Gamma-Astro-Data-Formats (GADF,~\cite{gadf}) and can export these into the FITS-based data format defined therein.
Additionally, \texttt{pyirf} contains functionality to calculate flux sensitivity of gamma-ray instruments according to the requirements laid out for CTA and the optimization of event selection criteria to obtain the best flux sensitivity.

\section{\texttt{protopipe}}
\label{sec:protopipe}
\noindent
\texttt{protopipe} is a pipeline prototype for CTA based on the \texttt{ctapipe} and \texttt{pyirf} libraries.
It is distributed as a python package on the \texttt{PyPI} platform; the latest release at the time of writing is \texttt{0.4.0.post1} ~\cite{protopipe} .
Started as an independent project for image cleaning studies by the CTA Consortium group at CEA-Saclay/IRFU, it has been developed as an open-source package for the whole consortium since September 2019.
Since then, its development has been steered by the will to substitute the historical pipelines currently in use for the production of the official IRFs for CTA.
Such pipelines have been inherited from the VERITAS (EventDisplay~\cite{eventdisplay}) and MAGIC (MARS~\cite{mars}) experiments and adapted to the CTA scenario by their maintainers.
Even if they provide satisfactory results, they are not in line with the software requirements of CTA and not easily exploitable by the whole consortium.
The development of \texttt{protopipe} is strongly influenced by a step-by-step comparison with such pipelines,
which translates in a continuous code migration into the \texttt{ctapipe} and \texttt{pyirf} libraries (algorithms and support of additional analysis operations).

\texttt{protopipe} has been built around the two libraries described in this work by constantly trying to support their latest stable releases.
It also provides a module for multivariate analysis using supervised machine-learning techniques (\texttt{protopipe.mva}),
used to reconstruct energy and particle type of the events.

The pipeline provides four tools based on \texttt{ctapipe}, \texttt{protopipe.mva}, \texttt{ctapipe} and \texttt{pyirf} respectively.
Each tool is a python executable configurable via YAML-based configuration files.
\texttt{protopipe} also provides a way to launch the analysis on computing grids featuring the DIRAC interware~\footnote{\shorturl{dirac.readthedocs.io/en/latest/}}.
The tools can be launched on the grid thanks to an interface code developed separately from the main package and based on CTADIRAC\footnote{\shorturl{github.com/cta-observatory/CTADIRAC}}, a version of the DIRAC middleware customized for CTA.

\subsection{Description of the pipeline workflow}\noindent
A full dataset composed of simulated events from primary gammas, protons and electrons is split at the beginning of the analysis in sub-datasets.
Depending on the workflow of choice, a step of the pipeline will correspond to a tool being applied to one or more sub-datasets.
The currently tested workflow is the following:

\begin{itemize}[nosep]
    \item part of the gamma rays are used to train an energy reconstruction model,
    \item part of the gamma rays and part of the protons are used to train a particle classification model (making use also of the reconstructed energy),
    \item the remaining gamma rays and protons together with the full electron dataset are fully analyzed.
\end{itemize}

\begin{figure}[htbp]
\centering
\includegraphics[scale=0.40]{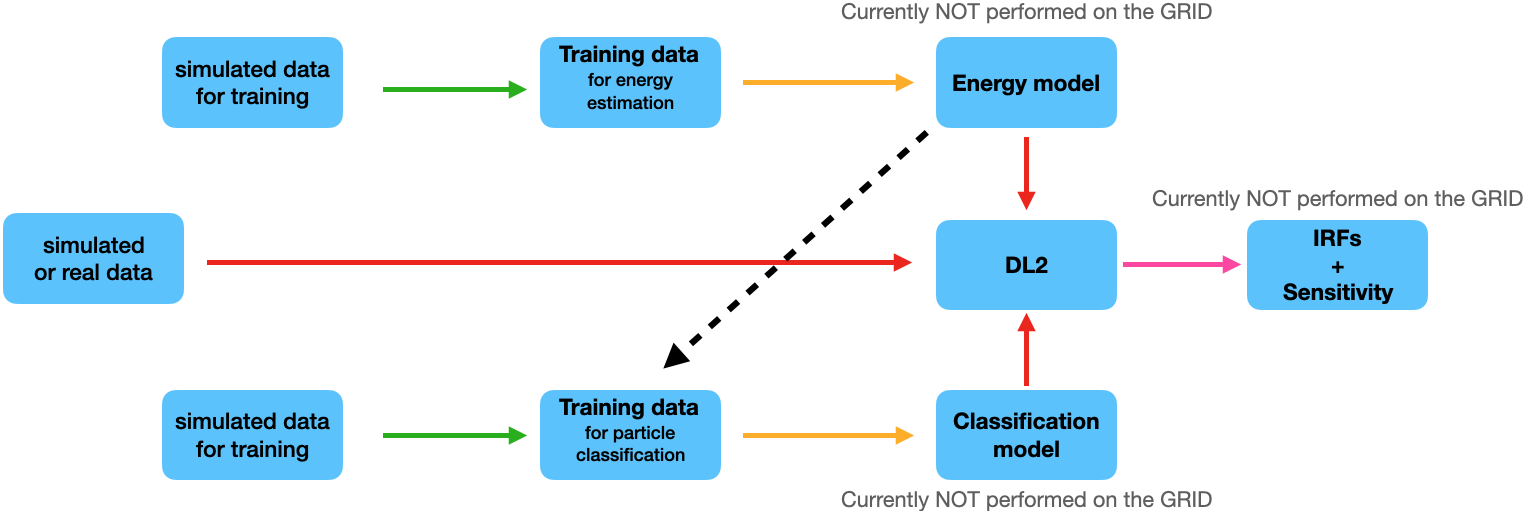}
\caption{Current pipeline workflow tested on full-scale analyses on the GRID. The actions performed by the tools 1, 2, 3 and 4 are highlighted by green, orange, red and pink arrows respectively. Reconstructed energy is used as a model feature when training particle classification (black dashed arrow)}
\label{fig:pipeline_workflow}
\end{figure}

In a real scenario of a gamma-ray analysis, the entire third sub-dataset and the proton sub-dataset used to train the particle classifier would correspond to data observed by the telescope array.
An overview of the workflow is shown in \autoref{fig:pipeline_workflow} and the tools are defined in the following sections.

\subsubsection{Tool 1: preparation of training data}\noindent
This tool is based on \texttt{ctapipe} and it produces data in a format suitable for model training.
This format is a combination of data levels as defined by the data models in \texttt{ctapipe}: DL1b (image parameters) and part of DL2 (reconstructed shower geometry).
The transformation of raw data into DL1b data makes use of the library capabilities described in paragraphs \ref{image_extraction} to \ref{image_parametrization}.
Since the pipeline workflow currently tested (see Fig.\ref{fig:pipeline_workflow}) comprises the use of two models (one for energy reconstruction and the other for particle classification) this tool is used in two separate steps of that analysis.

\subsubsection{Tool 2: production of models}\noindent
The production of machine-learning models is performed by the tool based on the \texttt{protopipe.mva} module. The dependencies are few: \texttt{numpy} and \texttt{pandas} to deal internally with tables of data, \texttt{joblib} for I/O support and \texttt{scikit-learn} to create the models and fit the test data.
The models currently tested are part of the \texttt{sklearn.ensemble} module: \texttt{AdaBoostRegressor} or \texttt{RandomForestRegressor} for energy reconstruction and \texttt{RandomForestClassifier} for particle classification.
It is possible to perform tuning of the hyper-parameters via an exhaustive search over lists of parameter values specified by the user (\texttt{sklearn.model\_selection.GridSearchCV}).
The tool outputs both model and tables of the events selected for training and testing as gzip-compressed pickled objects.

\subsubsection{Tool 3: production of fully-analyzed events}\noindent
This tool performs the full reconstruction pipeline and is applied to events which have to be independent from those used by the previous tools.
The operations performed are those described by paragraphs \ref{image_extraction} to \ref{estimation_events}.
In particular the tool requires as an input the models produced by Tool 2 in order to reconstruct both energy and particle type.
The models' input file format currently supported is the one output by the Tool 2.

\subsubsection{Tool 4: production of IRFs and optimized cuts}\noindent
This tool is based on the functions provided by the \texttt{pyirf} library and performs the following sequence of operations:
\begin{enumerate}[nosep]
    \item find the best cutoff in gammaness score, which is the result of the particle type classification, to best discriminate between signal and background, as well as the angular cut to obtain the best sensitivity for a given amount of observation time and a given template for the source of interest,
    \item estimate the sensitivity from the optimized cuts,
    \item compute the IRFs from the same selected events.
\end{enumerate}

The current output format is the one supported by \texttt{pyirf}: it builds on the data format specification given by the GADF integrated by input coming from CTA optimizations.

\section{Conclusions and Outlook}\noindent
\texttt{ctapipe} and \texttt{pyirf} offer open-source tools to solve a critical part
of the analysis of IACT data.
Using the IO plugin system, \texttt{ctapipe} can be used to process data by all experiments,
see for example \cite{magic-lst} for a combined analysis of LST-1 and MAGIC observations.
While the current version of \texttt{ctapipe} performs event property estimation using
geometrical or template based algorithms, the use of modern machine learning techniques
is also investigated (see for example \cite{gammalearn} and \cite{ctlearn}).
Performance of an analysis using \texttt{ctapipe} and \texttt{pyirf} on simulated data and first results on data from observations performed by the LST-1 are reported in~\cite{lst-performance}.
\texttt{protopipe} is being currently developed with the goal of superseding the reference analyses for the planned arrays, currently performed by EventDisplay and MARS.
It takes into account all supported cameras, optics and array configurations for CTA, enabling a high degree of flexibility to accommodate diverse instrument configurations.
It will be used to produce sets of IRFs sufficiently large to describe and investigate the performance of CTA under any required observing condition and science case and to analyze data from the whole set of telescopes.

\noindent\textbf{Acknowledgements:}
We gratefully acknowledge financial support from the agencies and organizations listed here:
\shorturl{www.cta-observatory.org/consortium_acknowledgments}

\renewcommand*{\bibfont}{\footnotesize}
\printbibliography

\section*{Full Authors List: CTA Consortium}
%
%
\scriptsize
\noindent
\input{authorlist.tex}
\end{document}

%% file: authorlist.tex
H.~Abdalla\textsuperscript{1}, H.~Abe\textsuperscript{2},
S.~Abe\textsuperscript{2}, A.~Abusleme\textsuperscript{3},
F.~Acero\textsuperscript{4}, A.~Acharyya\textsuperscript{5}, V.~Acín
Portella\textsuperscript{6}, K.~Ackley\textsuperscript{7},
R.~Adam\textsuperscript{8}, C.~Adams\textsuperscript{9}
% ...

\begin{enumerate}[nosep, noitemsep]
\item Centre for Space Research, North-West University, Potchefstroom, 2520, South Africa
\item Institute for Cosmic Ray Research, University of Tokyo, 5-1-5, Kashiwa-no-ha, Kashiwa, Chiba 277-8582, Japan
\item Pontificia Universidad Católica de Chile, Av. Libertador Bernardo O'Higgins 340, Santiago, Chile
\item AIM, CEA, CNRS, Université Paris-Saclay, Université Paris Diderot, Sorbonne Paris Cité, CEA Paris-Saclay, IRFU/DAp, Bat 709, Orme des Merisiers, 91191 Gif-sur-Yvette, France
\item Centre for Advanced Instrumentation, Dept. of Physics, Durham University, South Road, Durham DH1 3LE, United Kingdom
\item Port d'Informació Científica, Edifici D, Carrer de l'Albareda, 08193 Bellaterrra (Cerdanyola del Vallès), Spain
\item School of Physics and Astronomy, Monash University, Melbourne, Victoria 3800, Australia
\item Laboratoire Leprince-Ringuet, École Polytechnique (UMR 7638, CNRS/IN2P3, Institut Polytechnique de Paris), 91128 Palaiseau, France
\item Department of Physics, Columbia University, 538 West 120th Street, New York, NY 10027, USA
\item University of Oslo, Department of Physics, Sem Saelandsvei 24 - PO Box 1048 Blindern, N-0316 Oslo, Norway
\item EMFTEL department and IPARCOS, Universidad Complutense de Madrid, 28040 Madrid, Spain
\item Instituto de Astrofísica de Andalucía-CSIC, Glorieta de la Astronomía s/n, 18008, Granada, Spain
\item Institute of Space Sciences (ICE-CSIC), and Institut d'Estudis Espacials de Catalunya (IEEC), and Institució Catalana de Recerca I Estudis Avançats (ICREA), Campus UAB, Carrer de Can Magrans, s/n 08193 Cerdanyola del Vallés, Spain
\item Instituto de Física Teórica UAM/CSIC and Departamento de Física Teórica, Universidad Autónoma de Madrid, c/ Nicolás Cabrera 13-15, Campus de Cantoblanco UAM, 28049 Madrid, Spain
\item Dublin Institute for Advanced Studies, 31 Fitzwilliam Place, Dublin 2, Ireland
\item Universidad Nacional Autónoma de México, Delegación Coyoacán, 04510 Ciudad de México, Mexico
\item University of Geneva - Département de physique nucléaire et corpusculaire, 24 rue du Général-Dufour, 1211 Genève 4, Switzerland
\item INFN Dipartimento di Scienze Fisiche e Chimiche - Università degli Studi dell'Aquila and Gran Sasso Science Institute, Via Vetoio 1, Viale Crispi 7, 67100 L'Aquila, Italy
\item Instituto de Astronomia, Geofísico, e Ciências Atmosféricas - Universidade de São Paulo, Cidade Universitária, R. do Matão, 1226, CEP 05508-090, São Paulo, SP, Brazil
\item LUTH, GEPI and LERMA, Observatoire de Paris, CNRS, PSL University, 5 place Jules Janssen, 92190, Meudon, France
\item INAF - Osservatorio di Astrofisica e Scienza dello spazio di Bologna, Via Piero Gobetti 93/3, 40129 Bologna, Italy
\item INAF - Osservatorio Astrofisico di Arcetri, Largo E. Fermi, 5 - 50125 Firenze, Italy
\item INFN Sezione di Perugia and Università degli Studi di Perugia, Via A. Pascoli, 06123 Perugia, Italy
\item INFN Sezione di Napoli, Via Cintia, ed. G, 80126 Napoli, Italy
\item INFN Sezione di Roma Tor Vergata, Via della Ricerca Scientifica 1, 00133 Rome, Italy
\item Argonne National Laboratory, 9700 S. Cass Avenue, Argonne, IL 60439, USA
\item Aix-Marseille Université, CNRS/IN2P3, CPPM, 163 Avenue de Luminy, 13288 Marseille cedex 09, France
\item INAF - Osservatorio Astronomico di Roma, Via di Frascati 33, 00040, Monteporzio Catone, Italy
\item INAF - Osservatorio Astrofisico di Catania, Via S. Sofia, 78, 95123 Catania, Italy
\item Grupo de Electronica, Universidad Complutense de Madrid, Av. Complutense s/n, 28040 Madrid, Spain
\item National Astronomical Research Institute of Thailand, 191 Huay Kaew Rd., Suthep, Muang, Chiang Mai, 50200, Thailand
\item Instituto de Astrofísica de Canarias and Departamento de Astrofísica, Universidad de La Laguna, La Laguna, Tenerife, Spain
\item FZU - Institute of Physics of the Czech Academy of Sciences, Na Slovance 1999/2, 182 21 Praha 8, Czech Republic
\item Astronomical Institute of the Czech Academy of Sciences, Bocni II 1401 - 14100 Prague, Czech Republic
\item CCTVal, Universidad Técnica Federico Santa María, Avenida España 1680, Valparaíso, Chile
\item ETH Zurich, Institute for Particle Physics, Schafmattstr. 20, CH-8093 Zurich, Switzerland
\item The University of Manitoba, Dept of Physics and Astronomy, Winnipeg, Manitoba R3T 2N2, Canada
\item Department of Astronomy, University of Geneva, Chemin d'Ecogia 16, CH-1290 Versoix, Switzerland
\item Laboratoire Univers et Particules de Montpellier, Université de Montpellier, CNRS/IN2P3, CC 72, Place Eugène Bataillon, F-34095 Montpellier Cedex 5, France
\item Centro Brasileiro de Pesquisas Físicas, Rua Xavier Sigaud 150, RJ 22290-180, Rio de Janeiro, Brazil
\item Institut de Fisica d'Altes Energies (IFAE), The Barcelona Institute of Science and Technology, Campus UAB, 08193 Bellaterra (Barcelona), Spain
\item University of Groningen, KVI - Center for Advanced Radiation Technology, Zernikelaan 25, 9747 AA Groningen, The Netherlands
\item School of Physics, University of New South Wales, Sydney NSW 2052, Australia
\item INAF - Osservatorio Astrofisico di Torino, Strada Osservatorio 20, 10025 Pino Torinese (TO), Italy
\item Univ. Savoie Mont Blanc, CNRS, Laboratoire d'Annecy de Physique des Particules - IN2P3, 74000 Annecy, France
\item Department of Physics, TU Dortmund University, Otto-Hahn-Str. 4, 44221 Dortmund, Germany
\item University of Zagreb, Faculty of electrical engineering and computing, Unska 3, 10000 Zagreb, Croatia
\item University of Namibia, Department of Physics, 340 Mandume Ndemufayo Ave., Pioneerspark, Windhoek, Namibia
\item Nicolaus Copernicus Astronomical Center, Polish Academy of Sciences, ul. Bartycka 18, 00-716 Warsaw, Poland
\item Universität Hamburg, Institut für Experimentalphysik, Luruper Chaussee 149, 22761 Hamburg, Germany
\item Graduate School of Science, University of Tokyo, 7-3-1 Hongo, Bunkyo-ku, Tokyo 113-0033, Japan
\item Deutsches Elektronen-Synchrotron, Platanenallee 6, 15738 Zeuthen, Germany
\item Max-Planck-Institut für Kernphysik, Saupfercheckweg 1, 69117 Heidelberg, Germany
\item RIKEN, Institute of Physical and Chemical Research, 2-1 Hirosawa, Wako, Saitama, 351-0198, Japan
\item INFN Sezione di Padova and Università degli Studi di Padova, Via Marzolo 8, 35131 Padova, Italy
\item Escuela Politécnica Superior de Jaén, Universidad de Jaén, Campus Las Lagunillas s/n, Edif. A3, 23071 Jaén, Spain
\item Department of Physics and Electrical Engineering, Linnaeus University, 351 95 Växjö, Sweden
\item University of the Witwatersrand, 1 Jan Smuts Avenue, Braamfontein, 2000 Johannesburg, South Africa
\item Institut für Theoretische Physik, Lehrstuhl IV: Plasma-Astroteilchenphysik, Ruhr-Universität Bochum, Universitätsstraße 150, 44801 Bochum, Germany
\item Faculty of Physics and Applied Computer Science, University of Lódź, ul. Pomorska 149-153, 90-236 Lódź, Poland
\item INAF - Istituto di Astrofisica Spaziale e Fisica Cosmica di Milano, Via A. Corti 12, 20133 Milano, Italy
\item INFN and Università degli Studi di Siena, Dipartimento di Scienze Fisiche, della Terra e dell'Ambiente (DSFTA), Sezione di Fisica, Via Roma 56, 53100 Siena, Italy
\item Center for Astrophysics | Harvard \& Smithsonian, 60 Garden St, Cambridge, MA 02180, USA
\item INFN Sezione di Torino, Via P. Giuria 1, 10125 Torino, Italy
\item Finnish Centre for Astronomy with ESO, University of Turku, Finland, FI-20014 University of Turku, Finland
\item Pidstryhach Institute for Applied Problems in Mechanics and Mathematics NASU, 3B Naukova Street, Lviv, 79060, Ukraine
\item Bhabha Atomic Research Centre, Trombay, Mumbai 400085, India
\item Center for Astrophysics and Cosmology, University of Nova Gorica, Vipavska 11c, 5270 Ajdovščina, Slovenia
\item Institut für Astronomie und Astrophysik, Universität Tübingen, Sand 1, 72076 Tübingen, Germany
\item Research School of Astronomy and Astrophysics, Australian National University, Canberra ACT 0200, Australia
\item Department of Physics and Astronomy, University of California, Los Angeles, CA 90095, USA
\item INFN Sezione di Bari and Politecnico di Bari, via Orabona 4, 70124 Bari, Italy
\item Laboratoire de Physique des 2 infinis, Irene Joliot-Curie,IN2P3/CNRS, Université Paris-Saclay, Université de Paris, 15 rue Georges Clemenceau, 91406 Orsay, Cedex, France
\item INFN Sezione di Pisa, Largo Pontecorvo 3, 56217 Pisa, Italy
\item IRFU/DEDIP, CEA, Université Paris-Saclay, Bat 141, 91191 Gif-sur-Yvette, France
\item INAF - Osservatorio Astronomico di Padova, Vicolo dell'Osservatorio 5, 35122 Padova, Italy
\item INAF - Osservatorio Astronomico di Palermo "G.S. Vaiana", Piazza del Parlamento 1, 90134 Palermo, Italy
\item School of Physics, University of Sydney, Sydney NSW 2006, Australia
\item Sorbonne Université, Université Paris Diderot, Sorbonne Paris Cité, CNRS/IN2P3, Laboratoire de Physique Nucléaire et de Hautes Energies, LPNHE, 4 Place Jussieu, F-75005 Paris, France
\item Instituto de Física de São Carlos, Universidade de São Paulo, Av. Trabalhador São-carlense, 400 - CEP 13566-590, São Carlos, SP, Brazil
\item Departament de Física Quàntica i Astrofísica, Institut de Ciències del Cosmos, Universitat de Barcelona, IEEC-UB, Martí i Franquès, 1, 08028, Barcelona, Spain
\item Department of Physics, Washington University, St. Louis, MO 63130, USA
\item Saha Institute of Nuclear Physics, Bidhannagar, Kolkata-700 064, India
\item INAF - Osservatorio Astronomico di Capodimonte, Via Salita Moiariello 16, 80131 Napoli, Italy
\item Université de Paris, CNRS, Astroparticule et Cosmologie, 10, rue Alice Domon et Léonie Duquet, 75013 Paris Cedex 13, France
\item Astronomy Department of Faculty of Physics, Sofia University, 5 James Bourchier Str., 1164 Sofia, Bulgaria
\item Institut de Recherche en Astrophysique et Planétologie, CNRS-INSU, Université Paul Sabatier, 9 avenue Colonel Roche, BP 44346, 31028 Toulouse Cedex 4, France
\item School of Physics and Astronomy, University of Minnesota, 116 Church Street S.E. Minneapolis, Minnesota 55455-0112, USA
\item IRFU, CEA, Université Paris-Saclay, Bât 141, 91191 Gif-sur-Yvette, France
\item INAF - Istituto di Radioastronomia, Via Gobetti 101, 40129 Bologna, Italy
\item INAF - Istituto di Astrofisica Spaziale e Fisica Cosmica di Palermo, Via U. La Malfa 153, 90146 Palermo, Italy
\item Astronomical Observatory, Department of Physics, University of Warsaw, Aleje Ujazdowskie 4, 00478 Warsaw, Poland
\item Armagh Observatory and Planetarium, College Hill, Armagh BT61 9DG, United Kingdom
\item INFN Sezione di Catania, Via S. Sofia 64, 95123 Catania, Italy
\item INAF - Osservatorio Astronomico di Brera, Via Brera 28, 20121 Milano, Italy
\item Kavli Institute for Particle Astrophysics and Cosmology, Department of Physics and SLAC National Accelerator Laboratory, Stanford University, 2575 Sand Hill Road, Menlo Park, CA 94025, USA
\item Universidade Cruzeiro do Sul, Núcleo de Astrofísica Teórica (NAT/UCS), Rua Galvão Bueno 8687, Bloco B, sala 16, Libertade 01506-000 - São Paulo, Brazil
\item Universidad de Valparaíso, Blanco 951, Valparaiso, Chile
\item INAF - Istituto di Astrofisica e Planetologia Spaziali (IAPS), Via del Fosso del Cavaliere 100, 00133 Roma, Italy
\item Lund Observatory, Lund University, Box 43, SE-22100 Lund, Sweden
\item The Henryk Niewodniczański Institute of Nuclear Physics, Polish Academy of Sciences, ul. Radzikowskiego 152, 31-342 Cracow, Poland
\item Escola de Engenharia de Lorena, Universidade de São Paulo, Área I - Estrada Municipal do Campinho, s/n°, CEP 12602-810, Pte. Nova, Lorena, Brazil
\item INFN Sezione di Trieste and Università degli Studi di Udine, Via delle Scienze 208, 33100 Udine, Italy
\item Palacky University Olomouc, Faculty of Science, RCPTM, 17. listopadu 1192/12, 771 46 Olomouc, Czech Republic
\item Max-Planck-Institut für Physik, Föhringer Ring 6, 80805 München, Germany
\item CENBG, Univ. Bordeaux, CNRS-IN2P3, UMR 5797, 19 Chemin du Solarium, CS 10120, F-33175 Gradignan Cedex, France
\item Dublin City University, Glasnevin, Dublin 9, Ireland
\item Dipartimento di Fisica - Universitá degli Studi di Torino, Via Pietro Giuria 1 - 10125 Torino, Italy
\item Tata Institute of Fundamental Research, Homi Bhabha Road, Colaba, Mumbai 400005, India
\item Universitá degli Studi di Napoli "Federico II" - Dipartimento di Fisica "E. Pancini", Complesso universitario di Monte Sant'Angelo, Via Cintia - 80126 Napoli, Italy
\item Oskar Klein Centre, Department of Physics, University of Stockholm, Albanova, SE-10691, Sweden
\item Yale University, Department of Physics and Astronomy, 260 Whitney Avenue, New Haven, CT 06520-8101, USA
\item CIEMAT, Avda. Complutense 40, 28040 Madrid, Spain
\item University of Oxford, Department of Physics, Denys Wilkinson Building, Keble Road, Oxford OX1 3RH, United Kingdom
\item School of Physics \& Astronomy, University of Southampton, University Road, Southampton SO17 1BJ, United Kingdom
\item Department of Physics and Technology, University of Bergen, Museplass 1, 5007 Bergen, Norway
\item Western Sydney University, Locked Bag 1797, Penrith, NSW 2751, Australia
\item School of Physical Sciences, University of Adelaide, Adelaide SA 5005, Australia
\item INFN Sezione di Roma La Sapienza, P.le Aldo Moro, 2 - 00185 Roma, Italy
\item INFN Sezione di Bari, via Orabona 4, 70126 Bari, Italy
\item University of Rijeka, Department of Physics, Radmile Matejcic 2, 51000 Rijeka, Croatia
\item Institute for Theoretical Physics and Astrophysics, Universität Würzburg, Campus Hubland Nord, Emil-Fischer-Str. 31, 97074 Würzburg, Germany
\item Universidade Federal Do Paraná - Setor Palotina, Departamento de Engenharias e Exatas, Rua Pioneiro, 2153, Jardim Dallas, CEP: 85950-000 Palotina, Paraná, Brazil
\item Dept. of Physics and Astronomy, University of Leicester, Leicester, LE1 7RH, United Kingdom
\item Univ. Grenoble Alpes, CNRS, IPAG, 414 rue de la Piscine, Domaine Universitaire, 38041 Grenoble Cedex 9, France
\item National Centre for nuclear research (Narodowe Centrum Badań Jądrowych), Ul. Andrzeja Sołtana7, 05-400 Otwock, Świerk, Poland
\item Enrico Fermi Institute, University of Chicago, 5640 South Ellis Avenue, Chicago, IL 60637, USA
\item Institut für Physik \& Astronomie, Universität Potsdam, Karl-Liebknecht-Strasse 24/25, 14476 Potsdam, Germany
\item Department of Physics and Astronomy, Iowa State University, Zaffarano Hall, Ames, IA 50011-3160, USA
\item School of Physics, Aristotle University, Thessaloniki, 54124 Thessaloniki, Greece
\item King's College London, Strand, London, WC2R 2LS, United Kingdom
\item Escola de Artes, Ciências e Humanidades, Universidade de São Paulo, Rua Arlindo Bettio, CEP 03828-000, 1000 São Paulo, Brazil
\item Dept. of Astronomy \& Astrophysics, Pennsylvania State University, University Park, PA 16802, USA
\item National Technical University of Athens, Department of Physics, Zografos 9, 15780 Athens, Greece
\item University of Wisconsin, Madison, 500 Lincoln Drive, Madison, WI, 53706, USA
\item Astronomical Observatory of Taras Shevchenko National University of Kyiv, 3 Observatorna Street, Kyiv, 04053, Ukraine
\item Department of Physics, Purdue University, West Lafayette, IN 47907, USA
\item Unitat de Física de les Radiacions, Departament de Física, and CERES-IEEC, Universitat Autònoma de Barcelona, Edifici C3, Campus UAB, 08193 Bellaterra, Spain
\item Institute for Space-Earth Environmental Research, Nagoya University, Chikusa-ku, Nagoya 464-8601, Japan
\item Department of Physical Science, Hiroshima University, Higashi-Hiroshima, Hiroshima 739-8526, Japan
\item Department of Physics, Nagoya University, Chikusa-ku, Nagoya, 464-8602, Japan
\item Friedrich-Alexander-Universit\"{a}t Erlangen-N\"{u}rnberg, Erlangen Centre for Astroparticle Physics (ECAP), Erwin-Rommel-Str. 1, 91058 Erlangen, Germany
\item Santa Cruz Institute for Particle Physics and Department of Physics, University of California, Santa Cruz, 1156 High Street, Santa Cruz, CA 95064, USA
\item IRFU / DIS, CEA, Université de Paris-Saclay, Bat 123, 91191 Gif-sur-Yvette, France
\item INFN Sezione di Trieste and Università degli Studi di Trieste, Via Valerio 2 I, 34127 Trieste, Italy
\item School of Physics \& Center for Relativistic Astrophysics, Georgia Institute of Technology, 837 State Street, Atlanta, Georgia, 30332-0430, USA
\item Alikhanyan National Science Laboratory, Yerevan Physics Institute, 2 Alikhanyan Brothers St., 0036, Yerevan, Armenia
\item INAF - Telescopio Nazionale Galileo, Roche de los Muchachos Astronomical Observatory, 38787 Garafia, TF, Italy
\item INFN Sezione di Bari and Università degli Studi di Bari, via Orabona 4, 70124 Bari, Italy
\item University of Split - FESB, R. Boskovica 32, 21 000 Split, Croatia
\item Universidad Andres Bello, República 252, Santiago, Chile
\item Academic Computer Centre CYFRONET AGH, ul. Nawojki 11, 30-950 Cracow, Poland
\item University of Liverpool, Oliver Lodge Laboratory, Liverpool L69 7ZE, United Kingdom
\item Department of Physics, Yamagata University, Yamagata, Yamagata 990-8560, Japan
\item Astronomy Department, Adler Planetarium and Astronomy Museum, Chicago, IL 60605, USA
\item Faculty of Management Information, Yamanashi-Gakuin University, Kofu, Yamanashi 400-8575, Japan
\item Department of Physics, Tokai University, 4-1-1, Kita-Kaname, Hiratsuka, Kanagawa 259-1292, Japan
\item Centre for Astrophysics Research, Science \& Technology Research Institute, University of Hertfordshire, College Lane, Hertfordshire AL10 9AB, United Kingdom
\item Cherenkov Telescope Array Observatory, Saupfercheckweg 1, 69117 Heidelberg, Germany
\item Tohoku University, Astronomical Institute, Aobaku, Sendai 980-8578, Japan
\item Department of Physics, Rikkyo University, 3-34-1 Nishi-Ikebukuro, Toshima-ku, Tokyo, Japan
\item Department of Physics and Astronomy and the Bartol Research Institute, University of Delaware, Newark, DE 19716, USA
\item Institut für Astro- und Teilchenphysik, Leopold-Franzens-Universität, Technikerstr. 25/8, 6020 Innsbruck, Austria
\item Department of Physics and Astronomy, University of Utah, Salt Lake City, UT 84112-0830, USA
\item IMAPP, Radboud University Nijmegen, P.O. Box 9010, 6500 GL Nijmegen, The Netherlands
\item Josip Juraj Strossmayer University of Osijek, Trg Ljudevita Gaja 6, 31000 Osijek, Croatia
\item Department of Earth and Space Science, Graduate School of Science, Osaka University, Toyonaka 560-0043, Japan
\item Yukawa Institute for Theoretical Physics, Kyoto University, Kyoto 606-8502, Japan
\item Astronomical Observatory, Jagiellonian University, ul. Orla 171, 30-244 Cracow, Poland
\item Landessternwarte, Zentrum für Astronomie der Universität Heidelberg, Königstuhl 12, 69117 Heidelberg, Germany
\item University of Alabama, Tuscaloosa, Department of Physics and Astronomy, Gallalee Hall, Box 870324 Tuscaloosa, AL 35487-0324, USA
\item Department of Physics, University of Bath, Claverton Down, Bath BA2 7AY, United Kingdom
\item University of Iowa, Department of Physics and Astronomy, Van Allen Hall, Iowa City, IA 52242, USA
\item Anton Pannekoek Institute/GRAPPA, University of Amsterdam, Science Park 904 1098 XH Amsterdam, The Netherlands
\item Faculty of Computer Science, Electronics and Telecommunications, AGH University of Science and Technology, Kraków, al. Mickiewicza 30, 30-059 Cracow, Poland
\item Faculty of Science, Ibaraki University, Mito, Ibaraki, 310-8512, Japan
\item Faculty of Science and Engineering, Waseda University, Shinjuku, Tokyo 169-8555, Japan
\item Institute of Astronomy, Faculty of Physics, Astronomy and Informatics, Nicolaus Copernicus University in Toruń, ul. Grudziądzka 5, 87-100 Toruń, Poland
\item Graduate School of Science and Engineering, Saitama University, 255 Simo-Ohkubo, Sakura-ku, Saitama city, Saitama 338-8570, Japan
\item Division of Physics and Astronomy, Graduate School of Science, Kyoto University, Sakyo-ku, Kyoto, 606-8502, Japan
\item Centre for Quantum Technologies, National University Singapore, Block S15, 3 Science Drive 2, Singapore 117543, Singapore
\item Institute of Particle and Nuclear Studies, KEK (High Energy Accelerator Research Organization), 1-1 Oho, Tsukuba, 305-0801, Japan
\item Department of Physics and Astronomy, University of Sheffield, Hounsfield Road, Sheffield S3 7RH, United Kingdom
\item Centro de Ciências Naturais e Humanas, Universidade Federal do ABC, Av. dos Estados, 5001, CEP: 09.210-580, Santo André - SP, Brazil
\item Dipartimento di Fisica e Astronomia, Sezione Astrofisica, Universitá di Catania, Via S. Sofia 78, I-95123 Catania, Italy
\item Department of Physics, Humboldt University Berlin, Newtonstr. 15, 12489 Berlin, Germany
\item Texas Tech University, 2500 Broadway, Lubbock, Texas 79409-1035, USA
\item University of Zielona Góra, ul. Licealna 9, 65-417 Zielona Góra, Poland
\item Institute for Nuclear Research and Nuclear Energy, Bulgarian Academy of Sciences, 72 boul. Tsarigradsko chaussee, 1784 Sofia, Bulgaria
\item University of Białystok, Faculty of Physics, ul. K. Ciołkowskiego 1L, 15-254 Białystok, Poland
\item Faculty of Physics, National and Kapodestrian University of Athens, Panepistimiopolis, 15771 Ilissia, Athens, Greece
\item Universidad de Chile, Av. Libertador Bernardo O'Higgins 1058, Santiago, Chile
\item Hiroshima Astrophysical Science Center, Hiroshima University, Higashi-Hiroshima, Hiroshima 739-8526, Japan
\item Department of Applied Physics, University of Miyazaki, 1-1 Gakuen Kibana-dai Nishi, Miyazaki, 889-2192, Japan
\item School of Allied Health Sciences, Kitasato University, Sagamihara, Kanagawa 228-8555, Japan
\item Departamento de Astronomía, Universidad de Concepción, Barrio Universitario S/N, Concepción, Chile
\item Charles University, Institute of Particle \& Nuclear Physics, V Holešovičkách 2, 180 00 Prague 8, Czech Republic
\item Astronomical Observatory of Ivan Franko National University of Lviv, 8 Kyryla i Mephodia Street, Lviv, 79005, Ukraine
\item Kobayashi-Maskawa Institute (KMI) for the Origin of Particles and the Universe, Nagoya University, Chikusa-ku, Nagoya 464-8602, Japan
\item Graduate School of Technology, Industrial and Social Sciences, Tokushima University, Tokushima 770-8506, Japan
\item Space Research Centre, Polish Academy of Sciences, ul. Bartycka 18A, 00-716 Warsaw, Poland
\item Instituto de Física - Universidade de São Paulo, Rua do Matão Travessa R Nr.187 CEP 05508-090 Cidade Universitária, São Paulo, Brazil
\item International Institute of Physics at the Federal University of Rio Grande do Norte, Campus Universitário, Lagoa Nova CEP 59078-970 Rio Grande do Norte, Brazil
\item University College Dublin, Belfield, Dublin 4, Ireland
\item Centre for Astro-Particle Physics (CAPP) and Department of Physics, University of Johannesburg, PO Box 524, Auckland Park 2006, South Africa
\item Departamento de Física, Facultad de Ciencias Básicas, Universidad Metropolitana de Ciencias de la Educación, Santiago, Chile
\item Núcleo de Formação de Professores - Universidade Federal de São Carlos, Rodovia Washington Luís, km 235 CEP 13565-905 - SP-310 São Carlos - São Paulo, Brazil
\item Physik-Institut, Universität Zürich, Winterthurerstrasse 190, 8057 Zürich, Switzerland
\item Department of Physical Sciences, Aoyama Gakuin University, Fuchinobe, Sagamihara, Kanagawa, 252-5258, Japan
\item University of the Free State, Nelson Mandela Avenue, Bloemfontein, 9300, South Africa
\item Faculty of Electronics and Information, Warsaw University of Technology, ul. Nowowiejska 15/19, 00-665 Warsaw, Poland
\item Rudjer Boskovic Institute, Bijenicka 54, 10 000 Zagreb, Croatia
\item Department of Physics, Konan University, Kobe, Hyogo, 658-8501, Japan
\item Kumamoto University, 2-39-1 Kurokami, Kumamoto, 860-8555, Japan
\item University School for Advanced Studies IUSS Pavia, Palazzo del Broletto, Piazza della Vittoria 15, 27100 Pavia, Italy
\item Aalto University, Otakaari 1, 00076 Aalto, Finland
\item Agenzia Spaziale Italiana (ASI), 00133 Roma, Italy
\item Observatoire de la Cote d'Azur, Boulevard de l'Observatoire CS34229, 06304 Nice Cedex 4, Franc
\end{enumerate}